\documentclass[twocolumn,prl,aps,superbib,tightenlines,floatfix]{revtex4}
\usepackage{amsfonts,amsmath,amssymb,amsthm,bm}
\usepackage{graphicx}

\begin{document}
\bibliographystyle{apsrev}
\title{Counting Statistics in Nanoscale Junctions.}
\author{Yu-Shen Liu }
\author{Yu-Chang Chen}
\email{yuchangchen@mail.nctu.edu.tw}
\affiliation{Department of Electrophysics, National Chiao Tung University, 1001 Ta Hush
Road, Hsinchu 30010, Taiwan}
\date{\today}
\begin{abstract}

We present first-principles calculations for the third moment of
the current in atomic-scale junctions. We calculate this quantity
in terms of the effective single-particle wave-functions obtained
self-consistently within the static density-functional theory.
As an example, we investigate the relations among the conductance, the second and third moments
of the current for carbon-atom chains of various lengths bridging two metal electrodes. We find that
the conductance, the second-order and the third-order Fano factors show odd-even oscillation
with the number of carbon atoms with the third-order Fano factor positively correlated
to the conductance.

\end{abstract}
\maketitle

Nanoscale electronics has generated a tremendous wave of scientific
interest in the past decade due to prospects of device-size reduction
offered by atomic-level control of certain physical properties~\cite{Aviram}.
In addition, it has spurred great interest in the fundamental
understanding of quantum transport~\cite{book}.
One of these fundamental questions relates to the moments of the current.
For instance, the second moment - shot noise - defines the quantum
fluctuations of the current at zero temperature due to the quantization of charge.
Shot noise reaches the classical limit $2eI$, where $e$ is the electron charge and $I$ is
the average current~\cite{Schottky}, when electrons
in a conductor drift in a completely uncorrelated way as described by
a Poissonian distribution of current events. On the other hand, in the presence
of a junction with a narrow constriction, electrons, shot noise can be
expressed as $S_{2}\propto \sum_{n}T_{n}\left( 1-T_{n}\right)$
in terms of the transmission probabilities of each eigen-channel $T_n$~\cite{Blanter}.
It is a powerful tool for the exploration of quantum statistics of
non-equilibrium electrons~\cite{Ruitenbeek1,Natelson,MRE,Agra,chen,Lagerqvist,chen1,Yao}
and may provide a means to explore also local temperature effects in nano-structures~\cite{chen1}.
In fact, it has been recently employed to
characterize the signature of molecules/atomic wires in junctions~\cite{Ruitenbeek1,Kiguchi,Natelson}.

The higher moments of the current (although more difficult to measure and calculate)
provide deeper insight into the statistics of charge dynamics, and are therefore more
refined tools to characterize the signature of molecules in junctions. However, no studies
have considered higher moments of the current in truly atomic-scale systems.
To address this issue, we have developed a theoretical approach that combined with
static density functional theory (DFT), which allows us to compute correlations up to the
third moment of the current in atomic junctions.

\begin{figure}
\includegraphics[width=7cm]{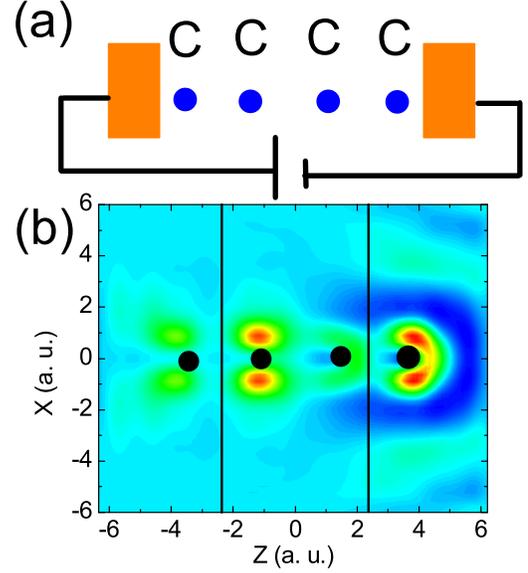}
\caption{ (color online) (a) Schematic of the four-carbon atomic junction.
(b) The spacial distribution of partial charge density for
electrons with energies near the Fermi levels shows $\pi$-orbital characters
at $V_{B}=0.01$~V.
Vertical black lines correspond to the edges of the jellium model and circles
correspond to atomic position.}
\label{Fig1}
\end{figure}

We have then investigated the relation between the conductance, the second moment
(or shot noise, $S_{2}$), and the third moment of the current (we denote it
with $S_{3}$) for a prototypical nanojunction consisting of an atomic chain with
different number of carbon atoms connecting
two metal electrodes as shown in Fig.~\ref{Fig1}(a). This is not just an academic example
since carbon is a versatile element capable of forming diverse structures
including diamond, graphite, fullerenes, nanotubes and graphene. Recently, experimentalists
have shown the possibility to form carbon atomic chains from graphite
using a transmission electron microscope~\cite{Jin}. The carbon atom chains have regularly
patterned electronic structures as a function of the number of carbon atoms~\cite{Avouris}.
In this regard, they are among
the few model systems in which theory and experiments can be reasonably compared.

\begin{figure}
\includegraphics[width=7cm]{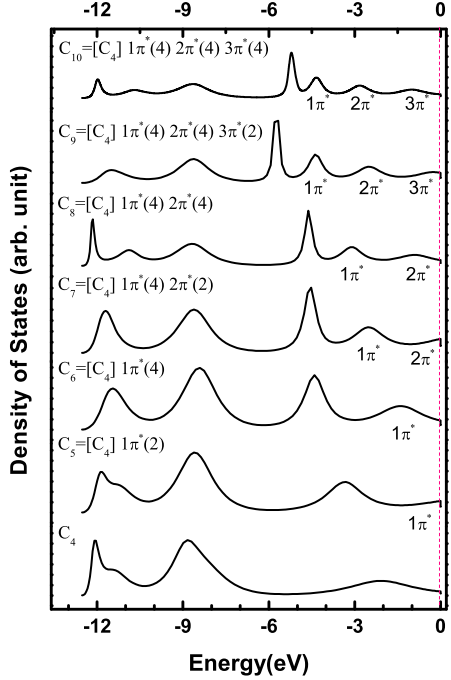}
\caption{ (color online) The density of states in the continuum region for N-atom
carbon chains sandwiched between two metal electrodes at $V_{B}=0.01$~V.
Zero of energy corresponds to the left Fermi level.}
\label{Fig2}
\end{figure}

Let us then consider a system at steady state subject to a bias $V_{B}=(E_{FR}-E_{FL})/e$,
where $E_{FR}$ and $E_{FL}$ are the right and left electrochemical potential, respectively.
The system is described by the field operator
\begin{equation}
\hat{\Psi}=\sum_{\alpha ,E,\mathbf{K}}a_{E\mathbf{K}}^{\alpha
}\left( t\right) \Psi _{E\mathbf{K}}^{\alpha }\left(
\mathbf{r}\right) , \label{field2}
\end{equation}%
where $\alpha=L$ or $R$; $a_{E\mathbf{K}}^{L(R)}\left( t\right)
=\exp (-i\omega t)a_{E\mathbf{K}}^{L(R)}$; and $a_{E\mathbf{K}}^{L(R)}$ is the
annihilation operators of electrons incident from the left (right) reservoir,
satisfying the anti-commutation relations
\begin{equation}
\{a_{E_{1}\mathbf{K}_{1}}^{\alpha },a_{E_{2}\mathbf{K}_{2}}^{\beta\dag}\}
=\delta _{\alpha \beta }\delta \left( E_{1}-E_{2}\right) \delta (\mathbf{K%
}_{1}-\mathbf{K}_{2}),  \label{anticom2}
\end{equation}%
where $\beta=L$ or $R$.  The single-particle wave functions
$\Psi_{E\textbf{K}}^{L(R)}$, describe electrons with energy E and
momentum $\textbf{K}$ incident from the left (L) and right (R) electrodes are later
computed self-consistently
in the framework of DFT~\cite{chen,DiVentra2,Lang}.

The current operator is defined as
\begin{eqnarray}
\hat{I}(z,t) &=&\frac{e\hbar }{mi}\sum_{E_{1}E_{2}}\sum_{\alpha \beta }\int d%
\mathbf{r\bot}\int d\mathbf{K}_{1}\int d\mathbf{K}_{2}  \notag \\
&&\cdot e^{i(E_{1}-E_{2})t/\hbar }a_{E_{1}\mathbf{K}_{1}}^{\alpha
\dag
}a_{E_{2}\mathbf{K}_{2}}^{\beta }\tilde{I}_{E_{1}\mathbf{K}_{1},E_{2}\mathbf{%
K}_{2}}^{\alpha \beta }(\mathbf{r}),  \label{currentop2}
\end{eqnarray}%
where
\begin{equation*}
\tilde{I}_{E_{1}\mathbf{K}_{1},E_{2}\mathbf{K}_{2}}^{\alpha \beta
}\left( \mathbf{r}\right) =\left( \Psi
_{E_{1}\mathbf{K}_{1}}^{\alpha }\right) ^{\ast }\nabla \Psi
_{E_{2}\mathbf{K}_{2}}^{\beta }-\nabla \left( \Psi
_{E_{1}\mathbf{K}_{1}}^{\alpha }\right) ^{\ast }\Psi _{E_{2}\mathbf{K}%
_{2}}^{\beta }.
\end{equation*}%

At zero temperature, the average of current operator gives
the first moment,
\begin{equation}
<\hat{I}>=\frac{e\hbar }{mi}\int_{E_{FL}}^{E_{FR}}dE\int d\mathbf{r_\bot}\int d%
\mathbf{K}\tilde{I}_{E\mathbf{K},E\mathbf{K}}^{R,R}\left(
\mathbf{r}\right), \label{currentavg2}
\end{equation}%
where the following expectation values have been used:
\begin{equation}
<a_{E_{1}\mathbf{K}_{1}}^{\alpha \dag
}a_{E_{2}\mathbf{K}_{2}}^{\beta
}>=\delta _{\alpha \beta }\delta \left( E_{1}-E_{2}\right) \delta (\mathbf{K}%
_{1}-\mathbf{K}_{2})f_{E}^{\alpha },  \label{expectation2}
\end{equation}%
with $f_{E}^{L\left( R\right) }$ the Fermi-Dirac distribution
function in the left (right) electrode.

We note that our wave-functions as obtained in the framework of DFT calculations describe
noninteracting electrons, where the effective single-particle
wavefunctions have boundary conditions describing electrons that are partially
transmitted and partially reflected. Unlike the formalism developed in, \emph{e.g.}, Ref.~\cite{Blanter}, which
has in-out ordering, the current operator defined in Eq.~(\ref{currentavg2})
describes steady-states current where the time ordering is not involved.
Therefore, the current correlation functions are defined in the time-unordered way~\footnote{This does not affect the second moment but
it is important for the third and higher moments.},
\begin{equation}
S_{2}(\omega)=2\pi \hbar \int d(t_{1}-t_{2})e^{i\omega(t_{1}-t_{2})}
<\Delta \hat{I}(t_{1}) \Delta \hat{I}(t_{2})>, \label{S2W}
\end{equation}
and
\begin{widetext}
\begin{equation}
S_{3}(\omega,\omega')=(2\pi \hbar)^{2} \int d(t_{1}-t_{3})
\int d(t_{2}-t_{3})e^{i\omega(t_{1}-t_{3})}e^{i\omega'(t_{2}-t_{3})}
<\Delta \hat{I}(t_{1}) \Delta \hat{I}(t_{2})\Delta \hat{I}(t_{3})>, \label{S3W}
\end{equation}
\end{widetext}
where $\triangle \hat{I}(t)=\hat{I}(t)-\left\langle \hat{I}\right\rangle$;
$S_{2}(\omega)$ and $S_{3}(\omega,\omega')$ are the two- and three-current
spectral functions, respectively.

The zero-frequency 2nd and 3rd moment of the steady-state current can
be defined as $S_{2}=S_{2}(\omega=0)$ and $S_{3}=S_{3}(\omega=0,\omega^{\prime}=0)$,
\begin{widetext}
\begin{equation}
S_{2}(z_{1},z_{2})=2\pi \hbar (\frac{e\hbar}{mi})^{2}\int_{E_{FL}}^{E_{FR}} dE \int d \textbf{r}_{1\bot}
\int d \textbf{r}_{2\bot} \int d \textbf{K}_{1}  \int d \textbf{K}_{2} \tilde{I}_{E \textbf{K}_{1},E\textbf{K}_{2}}^{RL}(\textbf{r}_{1})\tilde{I}_{E\textbf{K}_{2},
E \textbf{K}_{1}}^{LR}(\textbf{r}_{2}), \label{S2}
\end{equation}
\end{widetext}
and
\begin{widetext}
\begin{equation}
\begin{split}
S_{3}(z_{1},z_{2},z_{2})=&(2\pi
\hbar)^{2}(\frac{e\hbar}{mi})^{3}\int_{E_{FL}}^{E_{FR}} dE
\int d \textbf{r}_{1\bot} \int d \textbf{r}_{2\bot} \int d \textbf{r}_{3\bot}
\int d \textbf{K}_{1} \int d \textbf{K}_{2} \int d \textbf{K}_{3}\\&
[\tilde{I}_{E\textbf{K}_{1}E\textbf{K}_{2}}^{RL}(\textbf{r}_{1})
\tilde{I}_{E\textbf{K}_{2}E \textbf{K}_{3}}^{LL}(\textbf{r}_{2})
\tilde{I}_{E\textbf{K}_{3}E\textbf{K}_{1}}^{LR}(\textbf{r}_{3})
-
\tilde{I}_{E\textbf{K}_{1}E\textbf{K}_{3}}^{RL}(\textbf{r}_{1})
\tilde{I}_{E\textbf{K}_{2}E\textbf{K}_{1}}^{RR}(\textbf{r}_{2})
\tilde{I}_{E\textbf{K}_{3}E \textbf{K}_{2}}^{LR}(\textbf{r}_{3})].\label{S3}
\end{split}
\end{equation}
\end{widetext}
where the expectation values in Eqs.~(\ref{S2W}) and (\ref{S3W})
have been calculated using the Wick-Bloch-De Dominicis theorem~\cite{Kubo},
\begin{widetext}
\begin{equation}
<\hat{A}_{n}\hat{A}_{n-1}\cdots \hat{A}_{1}>=
\left\{
\begin{array}{c}
0  \text{, for }n=odd \text{,}
\\
\sum_{m=1}^{n-1} \eta^{n-m-1}<\hat{A}_{n} \hat{A}_{m}><\hat{A}_{n-1}\cdots \hat{A}_{m+1}\hat{A}_{m-1}\cdots \hat{A}_{1}>\text{, \ for }n=even\text{,}%
\end{array}%
\right.
\label{EqA}
\end{equation}%
\end{widetext}
where $\hat{A}_{i}$ denotes either creation or annihilation operators
and $\eta=-1 (1)$ for Fermions (Bosons). We note that alternative definitions
of Fourier transform, \emph{e.g.}, the integrals with respect to $(t_{1}-t_{2})$ and $(t_{2}-t_{3})$ in
Eq.~(\ref{S2W}) and (\ref{S3W}) may lead to different parametrization of
frequencies $\omega$ and $\omega'$ in the spectral functions. However, in the case of the steady-state current where $\omega=\omega'=0$,
the zero-frequency current correlations are independent of the choices of
Fourier transform. We also note that Eq.~(\ref{S2}) leads to the relations
$S_{2}^{RR}= S_{2}^{LL}=-S_{2}^{LR}=-S_{2}^{RL}$, where
$S_{2}^{RR}=S_{2}(z\rightarrow\infty,z\rightarrow\infty)$,
$S_{2}^{LL}=S_{2}(z\rightarrow-\infty,z\rightarrow-\infty)$,
$S_{2}^{LR}=S_{2}(z\rightarrow-\infty,z\rightarrow\infty)$, and
$S_{2}^{RL}=S_{2}(z\rightarrow\infty,z\rightarrow-\infty)$, which are a consequence of current conservation.
Similarly, Eq.~(\ref{S3}) leads to
\begin{equation}
S_{3}^{R(R,L)R}= S_{3}^{L(R,L)L}=- S_{3}^{L(R,L)R}=- S_{3}^{R(R,L)L},
\end{equation}

For a single-channel tunnel junction with transmission probability $T$,
the first, second, and third moment of current are given by
$I \propto T$, $S_{2} \propto T(1-T)$, and $S_{3} \propto -2T^{2}(1-T)$, respectively.
The result of the unordered third moment is consistent with the results of
time-unordered three-current correlations derived by other groups~\cite{Salo,Bachmann}. Finally, we define the second- and third-order
Fano factors (which is dimensionless) in the small bias regime
for steady-state currents as $F_{2}=S_{2}/(2eI)$
and $F_{3}=S_{3}/[(2e)^{2}I]$, respectively. As a direct result, for the single-channel junction
$G \propto T$, $F_{2} \propto (1-T)$, and $F_{3} \propto -2T(1-T)$, respectively.

\begin{figure}
\includegraphics[width=8cm]{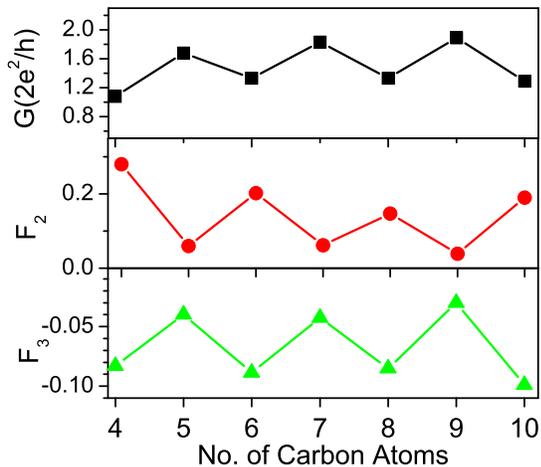}
\caption{ (color online) Conductance [(black) square; top panel], the second-order
Fano factor [(red) circle; middle panel], and
the third-order Fano factor [(green) triangle; bottom panel] of the atomic wires as a function of
the number of carbon atoms in the wire at $V_{B}=0.01$~V.}
\label{Fig3}
\end{figure}

As an example, we have investigated the counting statistics in linear atomic chains formed by
four to ten carbon atoms (denoted as C4 to C10) bridging between two metal
electrodes modeled as electron jellium ($r_{s}=2$). The distances between two
neighboring carbon atoms are $2.5$~a.~u., and the end atoms of the chain
are fixed at $1.4$~a.~u. inside the positive background edge of the electron
jellium as shown in Fig.~\ref{Fig1}(b)~\cite{Avouris}. As it was found in
Ref.~\cite{Avouris} when the length of the wire is increased
by one carbon atom, two electrons are added to the $\pi^{*}$-orbital as shown in Fig.~\ref{Fig2}.
The odd-numbered chains have a higher conductance due to
half-filled $\pi^{*}$-orbital while the even-numbered chains have a lower
conductance due to a full-filled $\pi^{*}$-orbital at the Fermi levels, as shown
in the upper panel of Fig.~\ref{Fig3}. In the middle and lower panels of
Fig.~\ref{Fig3} we show the influence of the number of carbon atoms on the
second-order $F_{2}$ and the third-order Fano factor $F_{3}$, respectively, in the linear
response regime ($V_{B}=0.01$~V). We observe
that $F_{2}$ and $F_{3}$ both display odd-even oscillation with the number of
carbon atoms.

\begin{figure}
\includegraphics[width=8cm]{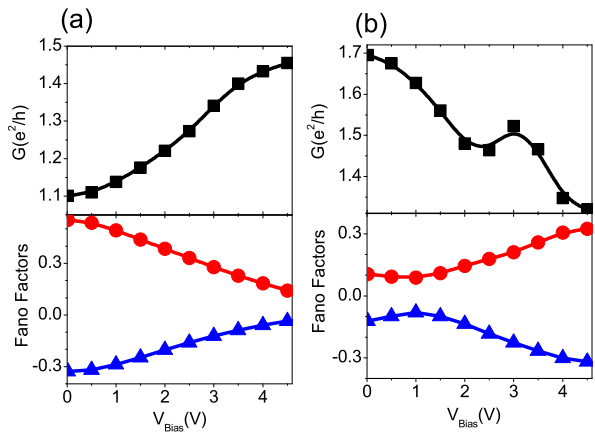}
\caption{ (color online) (a) For four-carbon atom wire and (b) for five-carbon atom wire:
the differential conductance [(black) square; top panel],
the second-order differential Fano factor [(red) circle; bottom panel], and
the three-order differential Fano factor [(blue) triangle; bottom panel] vs. bias.}
\label{Fig4}
\end{figure}

In order to better understand the relation among the moments of
counting statistics in the carbon atom chains, we have investigated the
differential conductance (defined as $G=\partial I/\partial V$),
the differential second-order Fano factor [defined as $F_{2}=(1/2e)\partial S_{2}/\partial I$],
and the differential third-order Fano factor [defined as $F_{3}=(1/2e)^{2}\partial S_{3}/\partial I$]
for the C4 and C5 wires in the non-linear response regime.
We observe that the differential conductance of C4 chain increases as the
applied bias increases, while the differential conductance of C5 chain
decreases as the applied bias increases, as shown in the top panels of
Figure~\ref{Fig4}(a) and (b), respectively. The increase (decrease) of differential
conductance with the biases for the C4 (C5) chain is due to the half-filled (full-filled)
$\pi^{*}$ orbital at the Fermi level, where more (less) states are included in
the current-carrying energy window created by increasing biases.
The bottom panels of Fig.~\ref{Fig4}(a) and (b) show that
conductance and $F_{3}$ are strongly positively correlated indicating that the
dominant eigen-channels for counting statistics have transmission probabilities $T>0.5$.

In conclusion, we have calculated the third moment of the current within
DFT that allows the study of counting
statistics at the atomic level. As an example, we have investigated
the relation among conductance, second and third moments of the current
for carbon chains of different length.
In the linear response regime,
conductance, second and third moments show odd-even oscillations with the number of
carbon atoms, which is mainly due to the orderly patterned electronic structure of
carbon-atom chains. In the nonlinear regime, the conductance increases (decreases) as
bias increases in even- (odd-) numbered carbon atom chains. We observe that $F_{3}$
and differential conductance are significantly positively correlated, thus
showing that third-order Fano factor provides more information than the
second-order Fano factor regarding the transmission probabilities of eigen-channels.

The authors thank MOE ATU, NCHC, National Center for Theoretical Sciences(South),
and NSC (Taiwan) for support under Grants NSC 97-2112-M-009-011-MY3,
098-2811-M-009-021, and 97-2120-M-009-005 and M. Di Ventra for useful discussions,
and S.~D. Yao for drawing Fig.~1 and 2.

\end{document}